\documentclass[a4paper,aps,prb,twocolumn]{revtex4}
\usepackage[latin1]{inputenc}
\usepackage{amsmath}
\usepackage{txfonts}
\usepackage{bm}
\usepackage{tikz}
\usepackage{graphicx}
\newcommand{\ie}{i.~e.,} 
\newcommand{\eg}{e.~g.,}
\newcommand{\hc}{{\text{H.\ c.}}}  
\newcommand{\ket}[1]{|#1\rangle}
\newcommand{\bra}[1]{\langle#1|}
\newcommand{\matel}[3]{\langle#1|\,#2\,|#3\rangle} 

\newcommand{\gc}{{\mathcal G}_2} 
\newcommand{\gset}{G_{}^S}

\newcommand{\cd}{c_{d}} 
\newcommand{\cdd}{c_{d}^\dagger}

\newcommand{\ed}{\varepsilon_d}
\newcommand{\ecal}{\mathcal{E}}
\newcommand{\lp}{\bm{\left(}} 
\newcommand{\rp}{\bm{\right)}}
\newcommand{\ha}{H_A}

\newcommand{\mc}[1]{\mathcal{#1}}
\newcommand{\vw}{W}
\newcommand{\jw}{J_W}
\newcommand{\za}{\mathcal{Z}}
\newcommand{\mmn}[1]{M_{mn}^{\,(#1)}}
\newcommand{\emn}{\mathcal{E}_{mn}}
\newcommand{\smn}{\mathcal{S}_{mn}}
\newcommand{\psim}{\ket{\Psi_m}}
\newcommand{\psimatel}[2]{\langle{\Psi_m}|#1|{\Psi_{#2}}\rangle}
\newcommand{\psinmatel}[2]{\langle{\Psi_n}|#1|{\Psi_{#2}}\rangle}
\newcommand{\ee}{\mathrm{e}}
\newcommand{\hnu}{h_{\mu}}
\newcommand{\zpart}{\mathcal{Z}}
\newcommand{\zone}{(01)}
\newcommand{\gr}[1]{\mathbb{G}_{#1}}
\newcommand{\gd}{\mathbb{G}_d}
\newcommand{\gzero}[1]{\mathbb{G}^{(0)}_{#1}}
\newcommand{\gammaW}{\Gamma_{W}}
\begin{document}
\title{Universality and the zero-bias conductance of the single-electron transistor}
\author{M. Yoshida} \affiliation{Departamento de F\'{\i}sica,
  Instituto de Geoci\^{e}ncias
  e Ci\^encias Exatas,\\ Universidade Estadual Paulista, 13500, Rio
  Claro, SP, Brazil}
\author{A. C. Seridonio}
\altaffiliation{Present address: Instituto de Física\\ Universidade Federal
  Fluminense, Niter\'oi, 24210-346, RJ- Brazil}

\author{L.~N.~Oliveira} \affiliation{Departamento de F\'{\i}sica e
  Inform\'{a}tica, Instituto
  de F\'{\i}sica de S\~{a}o Carlos, \\ Universidade de S\~{a}o Paulo,
  369, S\~{a}o Carlos, SP, Brazil}

\begin{abstract}
  The thermal dependence of the electrical conductance of the
  single-electron transistor (SET) in the zero-bias Kondo regime is
  discussed.  An exact mapping to the universal curve for the
  symmetric Anderson model is established. Linear, the mapping is
  parametrized by the Kondo temperature and the charge in the Kondo
  cloud. Illustrative numerical renormalization-group results, in
  excellent agreement with the mapping, are presented.
\end{abstract}

\maketitle

Nearly five decades ago, Anderson conceived a Hamiltonian to describe
the interaction between a magnetic impurity and otherwise free
conduction electrons.\cite{An61:41} Once a daunting theoretical
challenge, the Anderson Hamiltonian yielded to an essentially exact
numerical diagonalization,\cite{KWW80:1003} followed by an exact
analytical diagonalization.\cite{AFL83:331,TW83:453} From these and
alternative approaches, physical properties were extracted, which
eased the interpretation of experimental data;\cite{GGK+98:5225}
theoretical results provided unifying views of apparently unrelated
phenomena;\cite{Wil82:1} and quantitative comparisons brought forth
novel perceptions.\cite{LWC+87:1232}

The last ten years were especially fruitful. Parallel advances in
scanning tunneling spectroscopy and in the fabrication of
nanostructured semiconductor devices enhanced the interest in
transport
properties.\cite{MCJ+98.567,WFF+00:2105,HKS01:156803,WDE03.01,%
  AYR03.81,KZS+05.18824,Crommie05.1501,FFA03:155301,RWH+06:196601,%
  SSI+06_096603,ZB06.035332,SIS08:155304} In both areas, numerous experimental
breakthroughs and theoretical analyses were reported, and the Anderson
Hamiltonian proved spectacularly successful in more than one
occasion.\cite{OG03.02,FJC+07:256601}

Notwithstanding the substantial volume of exact results, certain
aspects of the model remain obscure. Consider universality, a concept
important in its own right and by virtue of its diverse
applications. Universal relations serve as benchmarks checking the
accuracy of numerical data; as resources promoting the convergence of
theoretical findings; and as instruments bridging the gap between the
theorist's tablet and the laboratory logbook. The conditions under
which the Anderson model exhibits universal thermodynamical properties
were identified.\cite{KWW80:1003,AFL83:331,TW83:453} Although one
expects all properties of the model to be universal in the same
domain, few firm results for the dynamical and transport properties
can be found in libraries.\cite{BCP08:395} Costi's et al's early
effort showed that the transport coefficients for the symmetric
Anderson model are universal.\cite{CHZ94.19} For asymmetric
models---even ones that display universal thermodynamical
properties---, nonetheless, the universal curves fail to fit the
numerical data, the disagreement growing with the (particle-hole)
asymmetry.

Puzzled by such contrasts, we have conducted a systematic study of the
transport properties for the Anderson Hamiltonian. We combined
analytical and numerical-renormalization group (NRG) tools and paid
special attention to universality. In a preliminary
report,\cite{SYO2007} we have discussed an Anderson model for a
quantum dot {\em side-coupled} to a quantum wire, a device comprising
two conduction paths whose transport properties are marked by
interference.\cite{KAS+04:035319,SAK+05:066801,KSA+06:36,OAK+07:084706,Kat07:233201}
Notwithstanding the constructive or destructive effects, we have been
able to identify universal behavior throughout the {\em Kondo regime},
the parametrical domain favoring the formation of a magnetic moment at
the quantum dot and its progressive screening by the conduction
electrons as the temperature is lowered past the scale set by the {\em
  Kondo temperature} $T_K$. Specifically, we found the thermal
dependence of the conductance to map linearly onto a universal function
of the temperature $T$ scaled by the Kondo temperature $T_K$. The
mapping is itself universal, \ie\ it depends on a single
physical property, the ground-state phase shift $\delta$, into which
the contributions from all model parameters are lumped.

This report examines the alternative experimental set-up in which a
quantum dot or molecule, instead of side-coupled to, is {\em embedded}
in the conduction
path.\cite{GSM+98.156,GGK+98.5225,GGH+00:2188,LSB02:725,OG03.02,%
  YKC+04.266802,YKC+05.256803,KZS+05.18824} We show that the thermal
dependence of the conductance maps onto the same universal
function. Although linear, the mapping now depends explicitly on a
model parameter---an external potential applied to the conduction
electrons---and hence contrasts with the conclusion in our previous
report. This dependence accounts for distinctions between the
transport properties in the embedded and side-coupled arrangements. At
high temperatures, for instance, potentials appropriately applied to
the conduction electrons in the side-coupled geometry drive the
conductance from low values up to the ballistic limit $\gc=2e^2/h$. If
the quantum dot is embedded in the conduction path, by contrast, the
high-temperature conductance is pinned at low values and virtually
insensitive to potentials applied to the conduction electrons. Our
analysis shows that, in the embedded configuration, the screening
charge in the Kondo cloud parametrizes the mapping to the universal
conductance curve. Since that charge is always close to unity, the
mapping is never far from the identity, with maximum relative deviations around
20\%.

Our presentation focuses the mapping between the SET and the universal
conductances. As illustrations we will present the results of a few
Numerical Renormalization Group (NRG) runs. A discussion of the
numerics, a comprehensive survey of the Kondo regime, and the
comparison with the side-coupled geometry will be deferred to another
report.

The text is divided in five Sections, more technical aspects of the
analysis having been confined to the three
Appendices. Section~\ref{sec:model} defines the
model. Section~\ref{sec:anderson-model} derives an expression relating the
conductance to the spectral density of the quantum dot
level. Section~\ref{sec:universality} is dedicated to universality,
and Section~\ref{sec:fixed-points}, to the fixed points of the model
Hamiltonian and to an extension of Langreth's exact expression for the
ground-state spectral density. Section~\ref{sec:cross} then shows
that, in the Kondo regime, the thermal dependence of the conduction can be mapped onto the
symmetric-SET universal conductance. Finally, Section~\ref{sec:conclusions}
collects our conclusions.

\section{Single-electron transistor\label{sec:model}}

\begin{figure}[th!]
  \includegraphics[width=\columnwidth]{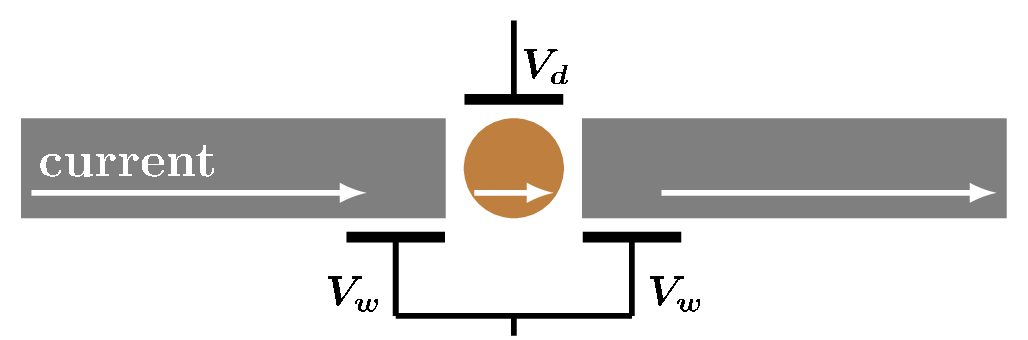}
  \caption{\label{fig:1}Single-electron transistor. A quantum dot (circle)
    bridges two quantum wires (rectangles). A gate potential $V_d$ controls
    the dot energy, while the symmetric potentials $V_w$ shift the 
    the energy of the wire orbitals close to the dot.}
\end{figure}

Figure~\ref{fig:1} depicts a single-electron transistor (SET), the
prototypical example of embedding.  The subject of numerous
experimental studies, the SET comprises two independent conduction
bands coupled by a localized level.

\begin{figure}[th]
  \includegraphics[width=\columnwidth]{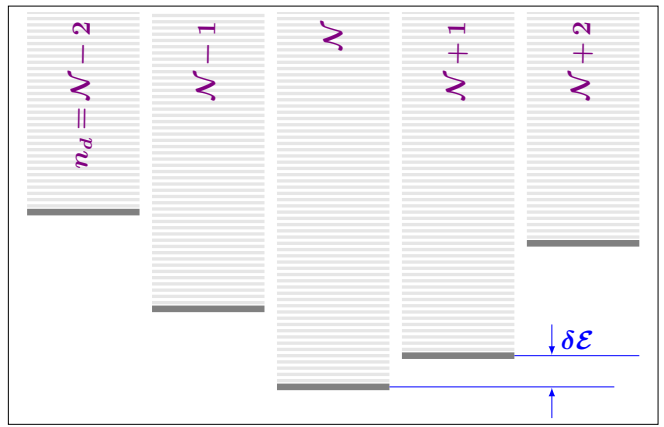}
  \caption{\label{fig:2}SET energies in the weak-coupling limit. The dot-level
    occupation $n_d$ labels the energies. For each $n_d$, the bold
    dash represents the conduction-band ground state, while the thinner
    lines represent excitations. The coupling between the dot and the
    two quantum wires mixes each level to the neighboring columns.}
\end{figure}

Qualitatively, the physics of Fig.~\ref{fig:1} was
understood long before the first device was
developed. Figure~\ref{fig:2} displays the spectrum of the SET
Hamiltonian $H$ for zero coupling. The dot levels being then decoupled
from the conduction bands, the eigenstates and eigenvalues of $H$ can
be labeled by the dot quantum numbers. For simplicity, we will
only refer to the dot occupation $n_d$. For fixed $n_d$, the
product of the lowest dot state by the conduction-band ground state is
shown as a bold dash. The gray levels above it represent the excited
states consistent with the same $n_d$ label.

A small transition amplitude $V$ between the quantum dot and the wires
is sufficient to modify this picture. The amplitude $V$ couples
strongly each gray level to the degenerate or nearly degenerate states in the
neighboring columns. Exceptions are the lowest levels in the column
labeled $n_d=\mc{N}$ in Fig.~\ref{fig:2}, which are energetically
distant from their neighbors and thus remain unperturbed to first
order in the coupling. At low temperatures, with $k_BT$ small in
comparison with the energy $\delta\mc{E}$ separating the ground state
from the closest level in the neighboring columns, the dot occupation
is frozen at $n_d=\mc{N}$, a constraint that raises the Coulomb
blockade against conduction through the dot.

Adjustment of the gate potential $V_d$ in Fig.~\ref{fig:1} brings
down the blockade. The potential shifts the energies of the dot levels
and can be tuned to the condition $\delta\mc{E}\approx0$, to impose
degeneracy between the bold dashes in the $n_d=\mc{N}$ and
$n_d=\mc{N}+1$ columns. An infinitesimal bias is then sufficient to
induce electronic flow between the wires throught the dot.  The
conductance peak, we see, at gate potentials $V_d$ such that the
ground-state expectation value of $n_d$ is half-integer, \eg\
$\matel{\Omega}{n_d}{\Omega}=\mc{N}+1/2$ as $\delta\mc{E}\to0$ in
Fig.~\ref{fig:2}.

Each peak identifies a resonance at the Fermi level.  As the gate
voltage is swept past $\delta\mc{E}=0$, the ground-state occupation
changes rapidly from $n_d=\mc{N}$ to $n_d=\mc{N}+1$, and as required
by the Friedel sum rule, so does the ground-state phase shift. At
moderately low temperatures, for thermal energies smaller than the
average spacing between the bold dashes in the figure, the plot of the
conductance as a function of the gate voltage is a succession of
peaks. Data collected in the laboratory at moderately low temperatures
do display a sequence of resonances. At very low temperatures,
however, the pattern changes to a sequence of intervals
alternating between insulation and conduction.

The conducting plateaus are due to the Kondo effect. For gate voltages
corresponding to odd ground-state dot occupations, the magnetic moment
of the resulting dot spin interacts antiferromagnetically with the
conduction electrons. As the device is cooled past the Kondo
temperature, the screening of the moment creates the {\em Kondo
  resonance}, a spiked enhancement of the density of states pinned at
the Fermi level. Notwithstanding the Coulomb blockade, the pinned
resonance allows conduction.

\section{Anderson model\label{sec:anderson-model}}

A variant of the Anderson Hamiltonian encapsulates the physics of the
device in Fig.~\ref{fig:1}. A spin degenerate level $\cd$ represents
the dot level, and two structureless half-filled conduction bands,
labeled $L$ (left) and $R$ (right), represent the two quantum
wires. The $L$ ($R$) wire comprises $N$ state $c_{kL}$ ($c_{kR}$)
with energies defined by the linear dispersion relation $\epsilon_k =
(k-k_F)v_F$ ($0\le k \le 2k_F$), so that the bandwidth is
$2D=2v_Fk_F$. The per-particle, per-spin density of conduction states
is $\rho=1/2D$, and we will let $\Delta\equiv D/N$ denote the energy
splittings in the conduction bands. The model Hamiltonian is then
the sum of three terms, $H=H_w+H_d+H_{wd}$, where the first term
describes the wires:
\begin{equation}
  \label{eq:hw}
  H_w =\sum_{k\alpha}\epsilon_k c_{k\alpha}^\dagger c_{k\alpha} 
  +\frac{\vw}{N}\sum_{kq\alpha}c_{k\alpha}^\dagger c_{q\alpha},
\end{equation}
with an intra-wire scattering potential $\vw$, fixed by the potential
$V_w$ in Fig.~\ref{fig:1}, and $\alpha=L, R$. The Hamiltonian $H_d$
describes the dot:
\begin{equation}
  \label{eq:dot}
  H_{d}=\ed n_d + Un_{d\uparrow}n_{d\downarrow},
\end{equation}
with a dot energy $\ed$ controlled by the gate potential $V_d$ in
Fig.~\ref{fig:1}; and the Hamiltonian $H_{wd}$ couples the wires to the dot:
\begin{equation}\label{eq:wd}
  H_{wd}= \frac{V}{\sqrt{2N}}\sum_{k\alpha} (c_{k\alpha}^\dagger\cd+\hc).
\end{equation}

\subsection{Parity \label{sec:parity}}

To exploit the inversion symmetry of Fig.~\ref{fig:1}, we define the
normalized even ($a_k$) and odd ($b_k$) operators
\begin{subequations}\label{eq:abk}
  \begin{eqnarray}\label{eq:ak}
    a_k = \frac1{\sqrt2}(c_{kL}+c_{kR});\\
    b_k = \frac1{\sqrt2}(c_{kL}-c_{kR}).\label{eq:bk}
  \end{eqnarray}
\end{subequations}

The projection of the model Hamiltonian on the basis of the $a_k$'s
and $b_k$'s splits it in two decoupled pieces, $H=\ha+H_B$, where
\begin{equation}\label{eq:ha}
  \ha=\sum_{k}\epsilon_k a_{k}^\dagger a_{k}
  +\vw f_0^\dagger f_0 +V(f_0^\dagger \cd+\hc)+H_{d},
\end{equation}
where we have introduced the traditional NRG shorthand
\begin{equation}\label{eq:f0}
  f_0\equiv\sum_k a_k/{\sqrt N},
\end{equation}
and 
\begin{equation}\label{eq:hb}
  H_B= \sum_{k}\epsilon_k b_{k}^\dagger b_{k} +\frac{\vw}N\sum_{kq}
  b_k^\dagger b_q.
\end{equation}

\subsection{Conductance\label{sec:cond}}

The odd Hamiltonian $H_B$ is decoupled from the quantum dot.  It is,
moreover, quadratic, and hence easily
diagonalizable. Appendix~\ref{sec:expr-cond} determines its spectrum,
analyzes the response of the conduction and dot electrons to the
application of an infinitesimal bias and turns the result into the
following Linear Response expression for the conductance:
\begin{equation}\label{eq:glin}
  G(T) = \gc\,\pi\gammaW\,\int_{-D}^{D}\rho_{d}(\epsilon,T) \left[-\frac{\partial
    f(\epsilon)}{\partial\epsilon}\right]\,d\epsilon,
\end{equation}
where $f(\epsilon)$ is the Fermi function;
\begin{equation}
  \label{eq:gammaW}
  \gammaW = \frac{\Gamma}{1+\pi^2\rho^2W^2}
\end{equation}
is the width $\Gamma=\pi\rho V^2$ of the $\cd$ level, here
renormalized by the scattering potential $W$; and
\begin{equation}
  \label{eq:rhod}
\rho_d(\epsilon,T)= \frac1{f(\epsilon)}\sum_{mn}\frac{e^{-\beta
    E_m}}{\za}|\matel{n}{\cdd}{m}|^2
\delta(\epsilon_{mn}-\epsilon)
\end{equation}
is the spectral density for the dot level. Here $\ket{m}$ and
$\ket{n}$ are eigenstates of $\ha$ with eigenvalues $E_m$ and $E_n$,
respectively, $\epsilon_{mn}\equiv E_m-E_n$, and $\za$ is the
partition function for the Hamiltonian $\ha$. 

As one would expect, given that the odd Hamiltonian $H_B$ commutes
with $\cd$, only the eigenvalues and eigenvectors of $\ha$ are needed
to compute the right-hand side of Eqs.~(\ref{eq:glin})~and
(\ref{eq:rhod}). The following discussion will hence focus the even
Hamiltonian, Eq.~(\ref{eq:ha}), which is equivalent to the
conventional spin-degenerate Anderson Hamiltonian.\cite{An61:41}

\subsection{Characteristic energies\label{sec:char-energ}}

\begin{figure}[th]
  \includegraphics[width=0.99\columnwidth]{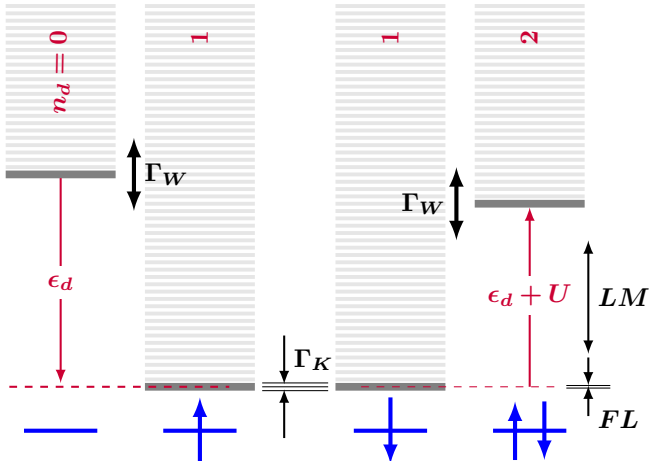}
  
  \caption{Spectrum of the spin-degenerate Anderson model, displayed
    as in Fig.~\ref{fig:2}. In the weak-coupling limit, the
    eigenstates are labeled by the occupation $n_d$ and spin component
    of the dot configuration displayed at the bottom. For $V\ne0$,
    each level in the left and right columns hybridizes with nearly
    degenerate levels in the central columns and acquires the width
    $\gammaW$ in Eq.~(\ref{eq:gammaW}). At low energies, the levels
    in the two central columns combine into a singlet and acquire a
    width $\Gamma_K\sim k_BT_K$.The vertical arrows near the right
    border mark the domains of the LM and FL fixed
    points.\label{fig:3} }
\end{figure}

Four characteristic energies govern the physical properties of the
Anderson Hamiltonian. Two of them are displayed in Fig.~\ref{fig:3}:
the energy $-\epsilon_d$ needed to remove an electron from the dot
level; and the energy $\epsilon_d+U$ needed to add an electron to the
level. The particle-hole transformation $c_k\to
c_k^\dagger$; $c_d\to -c_d^\dagger$ swaps the two energies, so that,
the transformed dot Hamiltonian is given by the right-hand side of
Eq.~(\ref{eq:dot}) with $\epsilon_d\to -(\epsilon_d+U)$.

If $2\epsilon_d+U=0$, the dot Hamiltonian remains invariant under the
particle-hole transformation. If, in addition, $W=0$,
Eq.~(\ref{eq:ha}) reduces to the symmetric Hamiltonian
\begin{equation}\label{eq:hasym}
  H_A^S= \sum_{k}\epsilon_k a_{k}^\dagger a_{k}
  +V(f_0^\dagger \cd+\hc)-\frac{U}2(n_{d\uparrow}-n_{d\downarrow})^2.
\end{equation}

With $V\ne0$, two other energies arise: the level width
$\gammaW$ [Eq.~(\ref{eq:gammaW})] and the Kondo energy $k_BT_K$,
given by
\begin{equation}  \label{eq:tk}
  T_K\sim \sqrt{\rho J}\exp(-1/\rho J),
\end{equation}
where $J$ is the antiferromagnetic interaction between the conduction
electrons and the dot magnetic moment,\cite{SW66:491}
\begin{equation}\label{eq:jk}
  \rho J=2\frac{\gammaW}{\pi|\ed|}\frac{U}{\ed+U}.
\end{equation}

In the Kondo regime, thermal and excitation energies are much smaller
than $\min(|\ed|, \ed+U)$. In Fig.~\ref{fig:3}, only the lowest levels
in the central columns are energetically accessible. The energy
$\gammaW$, associated with transitions from the central to the
external columns in the figure (\ie\ with $c_d^1\to c_d^2$ and
$c_d^1\to c_d^0$ transitions) becomes inoperant. Instead, at very low
excitation and thermal energies, smaller than the Kondo energy
$k_BT_K$, the dot spin binds antiferromagnetically to the conduction
spins. In Fig.~\ref{fig:3}, the lowest states in the left and right
central columns hybridize to constitute a Kondo singlet.

\section{Universality\label{sec:universality}}

\begin{figure}[th!]
  \includegraphics[width=0.99\columnwidth]{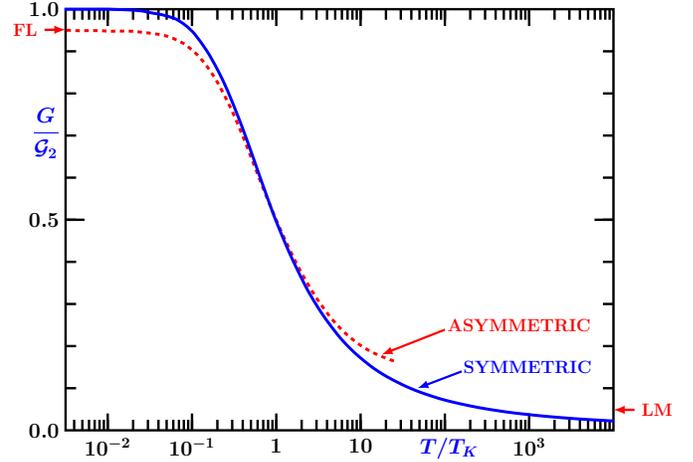}
  \caption[\ ]{\label{fig:4}Thermal dependences of the conductance for two sets of
    model paramters, obtained from Eqs.~(\ref{eq:glin})~and
    (\ref{eq:rhod}). The solid line depicts the universal conductance
    curve\cite{CHZ94.19} for the symmetric Hamiltonian~(\ref{eq:hasym})
    . Here, it was computed with $\Gamma= 0.1\,D$ and
    $U=3\,D$. The temperatures were scaled by the Kondo temperature
    $T_K=2.4\times10^{-6}\,D/k_B$, fixed by the requirement
    $G(T_K)=0.5\gc$. The dashed curve is the conductance for the
    Hamiltonian~(\ref{eq:ha}) with $\Gamma=0.1\,D$, $U=3\,D$,
    $\ed=-0.3\,D$, and $W=0$, which yielded
    $T_K=4.0\times10^{-3}\,D$.  To keep the data within the
    temperature range $k_BT<0.1\,D$, the dashed plot stops at
    $T=25\,T_K$. The horizontal arrows pointing to the
    vertical axes indicate the corresponding fixed-point conductances,
    given by Eqs.~(\ref{eq:glm})~and (\ref{eq:gfl}).}
\end{figure}

The concepts recapitulated in Section~\ref{sec:char-energ} emerged
over three decades ago, with the first accurate computation of the
magnetic susceptibility of the Anderson model,\cite{KWW80:1003} long
before the first essentially exact computation of the conductance. A
particularly important result in Costi's, Hewson's, and Zlatic's
survey of transport properties\cite{CHZ94.19} is the the thermal
dependence of the conductance for the symmetric Hamiltonian $H_A^S$,
the universal curve $\gset(T/T_K)$, depicted by the solid line in
Fig.~\ref{fig:4}. For $k_BT\ll D$ and any pair $(\Gamma, U)$
satisfying $\Gamma\ll U$ in Eq.~(\ref{eq:hasym}), proper adjustment of
the Kondo temperature $T_K$ gives a conductance curve $G(T/T_K)$ that
reproduces $\gset(T/T_K)$.

In Fig.~\ref{fig:4}, for instance, the solid line was computed from
the eigenvalues and eigenvectors of $H_A^S$ with $\Gamma=0.1\,D$ and
$U=3\,D$. The definition $G(T_K)\equiv0.5\gc$ yielded the Kondo
temperature $T_K=2.4\times10^{-6}\,D$. When the calculation was
repeated for $U=0.6\,D$ and the same $\Gamma$, the Kondo temperature
grew four orders of magnitude, to $T_K=2.2\times10^{-2}\,D$. Still,
for $k_BT<0.1\,D$, the plot of $G(T/T_K)$ resulted indistinguishable
from the solid curve. While $T_K$ is model-parameter dependent,
$G(T/T_K)$ is not.

Particle-hole asymmetry drives $G$ away from $\gset$.
For $U+2\epsilon_d\ne0$ or $W\ne0$, the universal curve $\gset(T/T_K)$
no longer matches $G(T/T_K)$. An example is the dashed curve in
Fig.~\ref{fig:4}, calculated with $\Gamma=0.1\,D$, $U=3\,D$,
$\ed=-0.3\,D$, and $W=0$. The definition $G(T_K)=0.5\gc$,
which in this case yields $T_K=4\times10^{-3}\,D$, forces the solid
and the dashed lines to agree at $T=T_K$; the conductance for the
asymmetric model nonetheless undershoots (overshoots) the universal
curve for $T < T_K$ ($T>T_K$). To reconcile this discrepancy with the
concept of universality, the following sections rely on
renormalization-group concepts.

\section{Fixed points\label{sec:fixed-points}}
Renormalization-goup theory probes the spectrum of Hamiltonians in
search of characteristic energies and scaling invariances. The wire
Hamiltonian~(\ref{eq:hw}), for instance, exhibits a single, trivial
characteristic energy: the conduction bandwidth $2D$. For energies
$\epsilon\ll D$, therefore, its spectrum is invariant under the
scaling transformation $H_w\to \Lambda H_w$, for arbitrary scaling
parameter $\Lambda>1$. Accordingly, for $\epsilon\ll D$, the wire
Hamiltonian is a stable {\em fixed point} of the renormalization-group
transformation in Ref.~\onlinecite{KWW80:1003}.

Latent in the Anderson Hamiltonian (\ref{eq:ha}), by contrast, are the
four nontrivial characteristic energies discussed in
Section~\ref{sec:char-energ}. Part of the spectrum of $\ha$ lies close
to fixed points; the remainder is in transition ranges. In the vicinity of
a fixed point, the spectrum remains approximately invariant under
scaling; in the transition intervals, the eigenvalues are comparable
to one or more characteristic energies and hence change rapidly under
scale transformations. In particular, the portion of the spectrum
pertinent to the Kondo regime comprises two lines of fixed points and
a crossover region.

For given thermal or excitation energy $\ecal$, the inequality
$\max(\ecal,\gammaW) \ll\min(|\ed|, \ed+U, D)$ defines the Kondo
regime. As Fig.~\ref{fig:3} shows, the dot occupation is then nearly
unitary. In the energy range $k_BT_K \ll \ecal \ll \min(|\ed|,\ed+U,
D)$, which is removed from characteristic energies, the Hamiltonian
$H_A$ is near the {\em Local Moment} fixed point (LM). At very low
energies, $\ecal \ll k_BT_K$, \ie\ below the energy scale defined by
the narrow set of levels at the center of Fig.~\ref{fig:3}, the
spectrum becomes asymptotically invariant under scaling as the
Hamiltonian approaches the Frozen Level fixed-point (FL). In the
intermediate region $\ecal\sim k_BT_K$, the Hamiltonian crosses over
from the LM to the FL.

\subsection{Fixed-point Hamiltonians\label{sec:fixed-point-hamilt}}
As the two central columns in Fig.~\ref{fig:3} indicate, the LM is
an unstable fixed-point consistent of a conduction band and a free
spin-1/2 variable. In the FL, a singlet replaces the spin, and the
Hamiltonian is equivalent to a conduction band---a stable fixed point.  
In their most general form, the fixed-point conduction bands mimic
the wire Hamiltonian, \ie\
\begin{equation}\label{eq:hlm}
    H_{LM}^*=\sum_k\epsilon_k a_k^\dagger a_k +W_{LM}f_0^\dagger f_0,
\end{equation}
and
\begin{equation}\label{eq:hfl}
    H_{FL}^*=\sum_k\epsilon_k a_k^\dagger a_k +W_{FL}f_0^\dagger f_0,
\end{equation}
with scattering potentials $W_{FL}$ and $W_{LM}$ dependent on $V$,
$W$, $U$ and $\epsilon_d$. Equations~(\ref{eq:hlm})~and (\ref{eq:hfl})
identify two lines of fixed points, parametrized by $W_{LM}$ and $W_{FL}$, respectively.

The Schrieffer-Wolff transformation offers an approximation for the LM potential:
\begin{equation}\label{eq:wSchW}
  \rho W_{LM} = \rho W +   2\frac{\gammaW}{\pi|\ed|}\frac{2\ed+U}{\ed+U}.
\end{equation}
For most applications, this expression is insufficiently accurate, and
an NRG computation is necessary to determine $W_{LM}$ and
$W_{FL}$. The exception is the Hamiltonian~(\ref{eq:hasym}), for which
$W_{LM}=0$, as required by particle-hole symmetry.

\subsection{Fixed-point phase shifts \label{sec:fixed-point-phase}}
Appendix~\ref{sec:diag} diagonalizes the quadratic
Hamiltonians~(\ref{eq:hlm})~and (\ref{eq:hfl}). For the LM, the
diagonal form reads
\begin{equation}\label{eq:hlmdiag}
  H_{LM}^*=\sum_k\varepsilon_\ell g_\ell^\dagger g_\ell,
\end{equation}
with phase-shifted energies
\begin{equation}\label{eq:erglm}
  \varepsilon_\ell = \epsilon_\ell -\frac{\delta_{LM}}{\pi}\Delta.
\end{equation}
At the LM, all conduction states are uniformly phase-shifted, with
\begin{equation}\label{eq:deltalm}
  \tan\delta_{LM} = -\pi\rho W_{LM}.
\end{equation}
For $\ha=H_A^S$, in particular, $\delta_{LM}=0$, and the low-energy eigenvalues
$\varepsilon_k$ coincide with the $\epsilon_k$.

The FL eigenvalues are likewise uniformly phase-shifted, 
\begin{equation}\label{eq:hfldiag}
  H_{FL}^*=\sum_k\tilde\varepsilon_k \tilde g_k^\dagger \tilde g_k,
\end{equation}
where $\tilde\varepsilon_k=\epsilon_k-(\delta/\pi)\Delta$. From
the Friedel sum rule, it follows that\cite{La66:516}
\begin{equation}\label{eq:deltafl}
  \delta = \delta_{LM}-\frac{\pi}2.
\end{equation}
For $\ha=H_A^S$, in particular, $\delta=\pi/2$.

\subsection{Conductance at the fixed points\label{sec:conductance-at-fixed}}
The LM is the fixed point to which the Anderson Hamiltonian would come
if $\Gamma=0$. For $0<\Gamma\ll \min(|\epsilon_d|, \epsilon_d+U, D)$,
although the renormalization-group flow never reaches the LM, it
brings $\ha$ close to the fixed point.  The substantial portion of the
spectrum of $\ha$ marked by the thin double-headed arrow in
Fig.~\ref{fig:3} is approximately described by the many-body
eigenvalues of $H_{LM}^*$, and in the pertinent energy range, the
physical properties of $\ha$ and $H_{LM}^*$ are approximately the
same. Likewise, at low temperatures, the properties of $\ha$ approach
those of $H_{FL}^*$.

The renormalization-group evolution of the Hamiltonian can
be traced in the termal dependence of the conductance. As the
temperature is reduced from $T\gg T_K$ to $T\ll T_K$, each curve in
Fig.~\ref{fig:4} crosses over from a lower plateau to a higher
one. The extension of Langreth's expression \cite{La66:516} derived in
Appendix~\ref{sec:extens-langr} determines the plateau conductances:
\begin{subequations}\label{eq:plateaus}
  \begin{eqnarray}\label{eq:glm}
    G_{LM}&=& \gc \sin^2(\delta_{LM} - \delta_W)
    =\gc\cos^2(\delta-\delta_W);\quad\\
    G_{FL}&=& \gc \sin^2(\delta - \delta_W),\label{eq:gfl}
  \end{eqnarray}
\end{subequations}
where $\delta_W$ is the ground-state phase shift for $V=0$.
According to the analysis in Appendix~\ref{sec:diag},
\begin{equation}
  \label{eq:tanDeltaW}
  \tan\delta_W = -\pi\rho W.
\end{equation}

The solid curve in Fig~\ref{fig:4} was computed for $\ha=H_A^S$, so
that $\delta_W=0$, while the ground-state (\ie\ FL) phase shift is
$\delta=\pi/2$. According to Eqs.~(\ref{eq:plateaus}), $G_{LM}=0$ and
$G_{FL}=\gc$, in agreement with the plot. The ground-state phase shift
for the dashed curve, extracted from the low-energy eigenvalues in the
NRG run that generated it, is somewhat lower: $\delta=0.43\pi$. Again
$\delta_W=0$, and the two horizontal arrows pointing to the vertical
axes in Fig.~\ref{fig:4} indicate the conductances predicted by
Eqs.~(\ref{eq:plateaus}). Given the relatively high Kondo temperature
($k_BT_K=4\times10^{-3}\,D$) in this run, the condition $k_BT \ll D$
restricts the curve to the range $T<25\,T_K$, so that, even at the
highest temperature shown, $\ha$ is relatively distant from the LM, and
Eq.~(\ref{eq:glm}) cannot be accurately checked. At low temperatures,
however, the renormalization-group flow bringing $\ha$ asymptotically
close to $H_{FL}^*$, the agreement with Eq.~(\ref{eq:gfl}) is
excellent.

\section{Crossover\label{sec:cross}}
In the Kondo regime, the Schrieffer-Wolff
transformation\cite{SW66:491} brings the Anderson Hamiltonian $\ha$ to
the Kondo form
\begin{equation}\label{eq:swkondo}
  H_J = \sum_k \epsilon_k a_k^\dagger a_k +W_{LM}f_0^\dagger f_0 + J
  \sum_{\mu\nu}f_{0\mu}^\dagger \bm{\sigma}_{\mu\nu}f_{0\nu}\cdot\bm{S},
\end{equation}
with $J$ defined in Eq.~(\ref{eq:jk}). 

To eliminate the scattering potential on the right-hand side, it is
convenient to project $H_J$ upon the basis of the eigenoperators $g_k$
of the LM, which yields\cite{irrelevant}
\begin{equation}\label{eq:lmkondo}
  H_J = \sum_k\varepsilon_\ell g_\ell^\dagger g_\ell +
  \jw\sum_{\mu\nu}\phi_{0\mu}^\dagger \bm{\sigma}_{\mu\nu}\phi_{0\nu}\cdot\bm{S},
\end{equation}
where $\jw=J\cos^2\delta_{LM}$, and
\begin{equation}\label{eq:phi0}
  \phi_0 = \frac1{\sqrt{N}}\sum_\ell g_\ell.
\end{equation}
In the symmetric case $\delta_{LM}$ vanishes, and the operator $\phi_0$ reduces to
$f_0$.

The second term on the right-hand side of Eq.~(\ref{eq:lmkondo})
drives the Hamiltonian from the LM to the FL. Along the resulting
trajectory, the eigenvalues of $H_J$ scale with
$T_K$.\cite{Wi75:773,KWW80:1044,TW83:453,AFL83:331} Let $T_{K}$ and
$\bar T_{K}<T_K$ be the Kondo temperatures correspondig to two sets
of model parameters in the Kondo regime:
$\mathcal{M}\equiv\left\{\Gamma,W, U, \ed\right\}$ and
$\bar{\mathcal{M}}\equiv\left\{\bar\Gamma,\bar W, \bar U,\bar
  \ed\right\}$, to which correspond the antiferromagnetic couplings
$J$ and $\bar J$, respectively.  If $\ket{m}$ is an eigenvector of
$H_{J}$ with eigenvalue $E_{m}$, then a corresponding eigenvector
$\ket{\bar m}$ of $H_{\bar J}$, the {\em scaling image of
  $\ket{m}$}, can always be found, with the same quantum numbers and
eigenvalue $\bar E_{m}$ such that $E_{m}/T_{K}=\bar E_{m}/\bar T_{K}$.

The matrix elements of any linear combination of the operators $g_k$
are moreover universal. Given two eigenstates $\ket{m}$ and $\ket{n}$
of $H_{J}$ and their scaling images $\ket{\bar m}$ and $\ket{\bar
  n}$, then the matrix elements of $\phi_0$, for example, are equal:
$\matel{m}{\phi_0}{n}=\matel{\bar m}{\phi_0}{\bar n}$.  Likewise,
the matrix elements of the operator
\begin{equation}
  \label{eq:phi1}
  \phi_1 = \sqrt{\frac3N}\sum_\ell\frac{\varepsilon_\ell}{D} g_\ell
\end{equation}
are universal: $\matel{m}{\phi_1}{n}=\matel{\bar m}{\phi_1}{\bar n}$.

\subsection{Thermal dependence of the conductance\label{sec:therm-depend}}
By contrast, the matrix elements $\matel{m}{\cd}{n}$ on the right-hand
side of Eq.~(\ref{eq:rhod}) are non-universal. Even at the lowest
energies, as Eq.~(\ref{eq:gamma_l0}) shows, they depends explicitly on
the model parameters. To discuss universal properties, therefore, we
must relate them to universal matrix elements, such as
$\matel{m}{\phi_0}{n}$, $\matel{m}{\phi_1}{n}$, or
$\matel{m}{g_\ell}{n}$. As a first step towards that goal, we evaluate
the commutator
\begin{equation}
  \label{eq:ha_comm_aq}
  [\ha, a_q^\dagger] = \epsilon_q a_q^\dagger +
  \frac{V}{\sqrt{N}}\cdd + \frac{W}{N}\sum_{p}a_p^\dagger,
\end{equation}
and sum the result over $q$, to find that
\begin{equation}
  \label{eq:ha_comm_f0}
  [\ha,f_0^\dagger] = \frac1{\sqrt3}f_1^\dagger + V\cdd
  + W f_0^\dagger.
\end{equation}
Here we have defined another shorthand
\begin{equation}
  \label{eq:f1}
  f_1=\sqrt{\frac3N}\sum_q \frac{\epsilon_q}D\, a_q.
\end{equation}

Equation~(\ref{eq:ha_comm_f0}) relates the matrix elements of $\cdd$
between two (low energy) eigenstates $\ket{m}$ and $\ket{n}$ of $\ha$
to those of the operators $f_0$ and $f_1$:
\begin{equation}
  \label{eq:aq_matel_cd}
  V\matel{m}{\cdd}{n}  = (E_m-E_n-W)\matel{m}{f_0^\dagger}{n}-\sqrt3
  D\matel{m}{f_1^\dagger}{n}.
\end{equation}
In the Kondo regime, with $\max(E_m,E_n)\ll D$, the first two
terms within the parentheses on the right-hand side can be dropped.

In the symmetric case, since $f_0$ ($f_1$) coincides with $\phi_0$
($\phi_1$), Eq.~(\ref{eq:aq_matel_cd}) shows that the product
$V\matel{m}{\cd}{n}$ is universal, in line with the firmly established
notion that $\Gamma\rho_d(\epsilon/k_BT_K, T/T_K)$, and $G^S(T/T_K)$
are universal functions.\cite{CHZ94.19,BCP08:395} To discuss
asymmetric Hamiltonians, we have to relate the operators $f_0$ and
$f_1$ to $\phi_0$ and $\phi_1$. This is done in
Appendix~\ref{sec:higher}, which shows that, in
the Kondo regime, a linear transformation with model-parameter
dependent coefficients relates the matrix elements
of both $f_0$ and $f_1$ to those of $\phi_0$ and $\phi_1$. When
Eq.~(\ref{eq:f0matelphi0}) is substituted for $f_0$ and $f_1$ on the
right-side of Eq.~(\ref{eq:aq_matel_cd}), it results that
\begin{equation}
  \label{eq:cd_phi01}
    \sqrt{\pi\rho\gammaW}\matel{m}{\cdd}{n}  =
    \alpha_0\matel{m}{\phi_0^\dagger}{n}+\alpha_1\matel{m}{\phi_1^\dagger}{n}.
\end{equation}
Here, the constants $\alpha_0$ and $\alpha_1$ are combinations of the
(unknown) linear coefficients on the right-hand side of
Eq.~(\ref{eq:f0matelphi0}), the parameter $W$ on the right-hand
side of Eq.~(\ref{eq:aq_matel_cd}), and the ratio $\sqrt{\pi\rho\gammaW}/V$,
by which we multiplied Eq.~(\ref{eq:aq_matel_cd}) to shorten the
following algebra.

Substitution in Eq.~(\ref{eq:rhod}) yields an expression relating
the spectral density $\rho_d$ to universal functions:
\begin{equation}
  \label{eq:cdUniversal}
\pi\rho\gammaW\rho_d(\epsilon,T) =
\alpha_0^2\rho_0(\epsilon,T)+\alpha_1^2\rho_1(\epsilon,T)
+\alpha_0\alpha_1\rho_{\zone}(\epsilon, T),
\end{equation}
where
\begin{eqnarray}\label{eq:rhos}
  \rho_{j}\,(\epsilon,T)&= \displaystyle\sum_{mn}&\frac{e^{-\beta
    E_m}}{\za f(\epsilon)}|\matel{n}{\phi_j}{m}|^2\nonumber\\
&&\times\,\delta(E_m-E_n-\epsilon)\qquad(j=0,1),
\end{eqnarray}
and
  \begin{eqnarray}
    \rho_{\zone}\,(\epsilon,T)&= \displaystyle\sum_{mn}&\frac{e^{-\beta E_m}}{\za
      f(\epsilon)}\,\left(\matel{m}{\phi_0^\dagger}{n}\matel{n}{\phi_1}{m}+\text{c.~c.}
    \right)\nonumber\\
    &&\times\,\delta(E_m-E_n-\epsilon)\label{eq:rho01}.
  \end{eqnarray}

Next, we substitute Eq.~(\ref{eq:cdUniversal}) on the right-hand side
of Eq.~(\ref{eq:glin}), to split the conduction into three pieces:
\begin{equation}
  \label{eq:g01bar}
    G(T) = \alpha_0^2 G_0(T) +\alpha_1^2G_1(T) 
    + \alpha_0\alpha_1 G_{\zone}(T),
\end{equation}
where
\begin{equation}\label{eq:g01}
  G_j(T)=\frac{\gc}{\rho}\,\int_{-D}^{D}\rho_{j}(\epsilon,T) \left[-\frac{\partial
    f(\epsilon)}{\partial\epsilon}\right]\,d\epsilon\qquad(j=0,1),
\end{equation}
and
\begin{equation}
\label{eq:g_0110}
G_{\zone}(T)=\frac{\gc}{\rho}\int_{-D}^{D}
\rho_{\zone}(\epsilon,T)
\left[-\frac{\partial f(\epsilon)}{\partial\epsilon}\right]\,d\epsilon. 
\end{equation}
\subsection{Universal contributions to the conductance\label{sec:univ-contrib}}
Given the universality of the energies $E_{m}$ and of the matrix
elements $\matel{m}{\phi_j}{n}$ ($j=1,2$) on the right-hand sides of
Eqs.~(\ref{eq:rhos})~and (\ref{eq:rho01}), we see that the spectral
densities $\rho_j(\epsilon,T)$ ($j=0,1$), and
$\rho_{\zone}(\epsilon,T)$ are universal. Inspection of the right-hand
sides of Eqs.~(\ref{eq:g01}-\ref{eq:g_0110}) shows that the functions
$G_j$ ($j=0,1$) and $G_{\zone}$ are likewise universal. To compute
them, we are free to consider any convenient Kondo-regime Hamiltonian.

Particle-hole symmetry makes $\ha^S$ especially convenient. To show
that the cross terms make no contribution to the conductance, \ie\
that $G_{\zone}(T)=0$, we only have to notice that, while
leaving $\ha^S$ unchanged, the particle-hole transformation
$\cd\to-\cdd$, $g_k\to g_k^\dagger$ (\ie\ $a_k\to a_k^\dagger$)
reverses the sign of the product of matrix elements
$\matel{m}{\phi_i^\dagger}{n}\matel{n}{\phi_j}{m}+\text{c.~c.}$ on the
right-hand side of Eq.~(\ref{eq:rho01}). We see that
$\rho^{\zone}(\epsilon, T)$ is an odd function of $\epsilon$, so that
the integral on the right-hand side of Eqs~(\ref{eq:g_0110}) vanishes.

To evaluate $G_0$ and $G_1$, we start out from the closed form
resulting from the diagrammatic expansion (in the coupling $V$) of the
conduction-electron retarded Green's function for the symmetric
Hamiltonian:
\begin{equation}
  \label{eq:gkk}
  \gr{kk'}^S(\epsilon) = \gzero{k}(\epsilon)\delta_{kk'} +
  \frac{V^2}N\gzero{k}(\epsilon)\gd^S(\epsilon)\gzero{k'},
\end{equation}
where $\gd^S$ is the retarded dot-level Green's function for the
symmetric Hamiltonian, and
\begin{equation}
  \label{eq:gzerok}
  \gzero{k}(\epsilon)=\frac{1}{\epsilon-\epsilon_k+i\eta}
\end{equation}
is the free conduction-electron retarded Green's function. 

From $\gr{kk'}^S$, it is a simple matter to obtain the spectral
densities on the right-hand side of Eq.~(\ref{eq:cdUniversal}): 
  \begin{equation}
    \label{eq:rho0gk}
    \rho_0(\epsilon,T) = -\frac1{\pi N}\Im\sum_{k k'}\gr{kk'}^S(\epsilon),
  \end{equation}
and
\begin{equation}
  \label{eq:rho1gk}
      \rho_1(\epsilon,T) = -\frac{3}{\pi ND^2}
      \Im\sum_{k k'}\epsilon_k\epsilon_{k'}\gr{kk'}^S(\epsilon).
\end{equation}

To compute the conductances at temperatures $T$ satisfying $k_BT\ll
D$, we only need the spectral densities for $\epsilon \ll D$. It is
appropriate, therefore, to expand the right-hand side of
Eq.~(\ref{eq:gzerok}) to linear order in $\epsilon/D$:
\begin{equation}
  \label{eq:gzeroklinear}
  \gzero{k}(\epsilon)=\frac{2\epsilon}{D}
  -i\pi\delta(\epsilon-\epsilon_k)\qquad(\epsilon\ll D).
\end{equation}

The sums over momenta on the right-hand side of
Eqs.~(\ref{eq:rho0gk})~and (\ref{eq:rho1gk}) are then easily
computed. Among the resulting terms, only the even powers of
$\epsilon$ contribute to the integral on the right-hand side of
Eq.~(\ref{eq:g01}). To compute the conductance to $\mc{O}[(k_BT/D)^2]$
we hence neglect the terms of
$\mc{O}(\epsilon/D)$. Equation~(\ref{eq:rho0gk}) then gives
\begin{equation}
  \label{eq:rho0eps}
  \rho_0(\epsilon, T) = \rho -{\pi\rho
    \Gamma}\rho_d^S(\epsilon,T),
\end{equation}
equivalent to an expression obtained in Ref.\cite{MNU04:3239}.

Substitution of this result for $\rho_0$ on the right-hand side of
Eq.~(\ref{eq:g01}) establishes a simple relation between the universal
function $G_0$ and the universal conduction for the symmetric
Hamiltonian:
\begin{equation}
  \label{eq:g0}
  G_0(T) = \gc - \gset(T).
\end{equation}

\begin{figure}[th]
  \includegraphics[width=0.99\columnwidth]{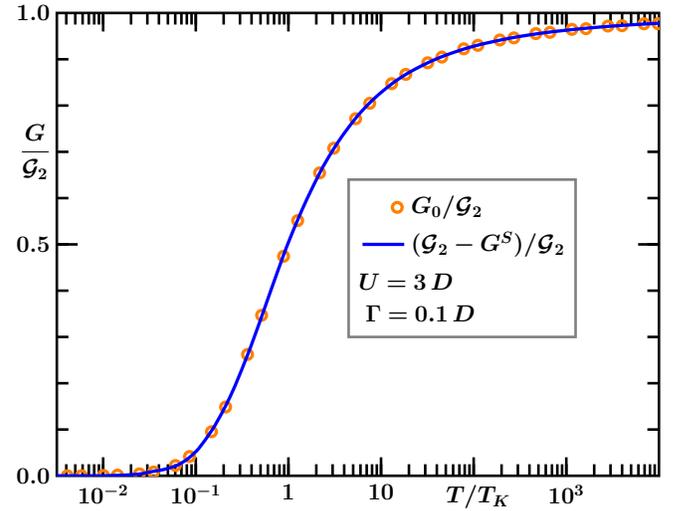}
  \caption{NRG results for the thermal dependence of the auxiliary
    conductance $G_0(T)$, associated with the spectral density for the
    operator $\phi_0$. The open circles show Eq.~(\ref{eq:g01}) for
    $j=0$, computed for the symmetric Hamiltonian with the displayed
    model parameters. The solid line is the right-hand side of
    Eq.~(\ref{eq:g0}), \ie\ the universal curve in
    Fig.~\ref{fig:4} subtracted from the quantum conductance $\gc$.}
  \label{fig:5}
\end{figure}

Equation~(\ref{eq:g0}) becomes exact, asymptotically, at low
temperatures. The deviations, of $\mc{O}[(k_BT/D)^2]$, are
insifignicant. As an illustration, the open circles in
Fig.~\ref{fig:5} show NRG data for the conductance $G_0(T)$,
Eq.~(\ref{eq:g01}), in excellent agreement with the solid line
representing the right-hand side of Eq.~(\ref{eq:g0}).

To the same accuracy, we can neglect the $\mc{O}(\epsilon/D)$ terms
resulting from the summation on the right-hand side of
Eq.~(\ref{eq:rho1gk}), which yields
\begin{equation}
  \label{eq:rho1eps}
  \rho_1(\epsilon, T) = \frac{6\rho \Gamma}{\pi}\rho_d^S(\epsilon,T).
\end{equation}
Equation~(\ref{eq:g01}) then shows that $G_1$ is also related to the
conductance for the symmetric Hamiltonian:
\begin{equation}
  \label{eq:g1}
  G_1(T)= \frac{6}{\pi^2}\gset(T),
\end{equation}

\subsection{Mapping to the universal conductance\label{sec:mapp-univ-cond}}
The combination of Eqs.~(\ref{eq:g0}) and~(\ref{eq:g1})
with the result $G_{\zone}=0$ reduces Eq.~(\ref{eq:g01bar}) to the equality
\begin{equation}
  \label{eq:ga0a1set}
  G(T) = \alpha_0^2\lp\gc -\gset(T)\rp
  +\alpha_1^2\,\frac6{\pi}\,\rho\,\gset(T)
\end{equation}

To determine the coefficients $\alpha_0$ and $\alpha_1$, we need only
compare the right-hand side with the fixed-point expressions for the
conductance. At the LM, $\gset=0$, and Eq.~(\ref{eq:glm}) shows
that $\alpha_0^2= \cos^2(\delta-\delta_W)$. At the FL, $\gset=\gc$,
and Eq.~(\ref{eq:gfl}) shows that
$(6/{\pi}^2)\alpha_1^2=\sin^2(\delta-\delta_W)$. These two results turn
Eq.~(\ref{eq:ga0a1set}) into the mapping
\begin{equation}\label{eq:guniversal}
  G\lp\frac{T}{T_K}\rp -\frac{\gc}2 = -
\lp\gset\lp\frac{T}{T_K}\rp-\frac{\gc}2\rp\cos2(\delta-\delta_W).
\end{equation}

\subsection{Illustrative numerical results\label{sec:illustration}}

Equation~(\ref{eq:tk}) offers an approximation for $T_K$, and
Eqs.~(\ref{eq:wSchW}), (\ref{eq:deltalm})~and (\ref{eq:deltafl})
provide an approximation for the ground-state phase shift $\delta$.
These estimates are far from the accuracy needed to fit numerical or
experimental data. In the laboratory, $T_K$ and $\delta-\delta_W$ are
adjustable parameters; the former, in particular, is determined by the
condition $G(T_K) = \gc/2$.\cite{GGK+98.5225,KSA+06:36,SAK+05:066801}
In the computer office, the two unknown parameters on the right-hand
side of Eq.~(\ref{eq:guniversal}) can can be extracted from the
conductance itself, or from other properties of the model
Hamiltonian. The phase shift $\delta$ is most easily obtained from the
ground-state eigenvalues of $\ha$. To determine the Kondo temperature
$T_K$, it has been traditional to fit the thermal dependence of the
magnetic susceptibility $\chi(T)$ with the universal curve for $k_B
T\chi(T/T_K).$\cite{KWW80:1003,TW83:453} Here, however, we prefer the
laboratory definition, which insures that both sides of
Eq.~(\ref{eq:guniversal}) vanish at $T=T_K$.

\begin{figure}[th]
  \includegraphics[width=0.99\columnwidth]{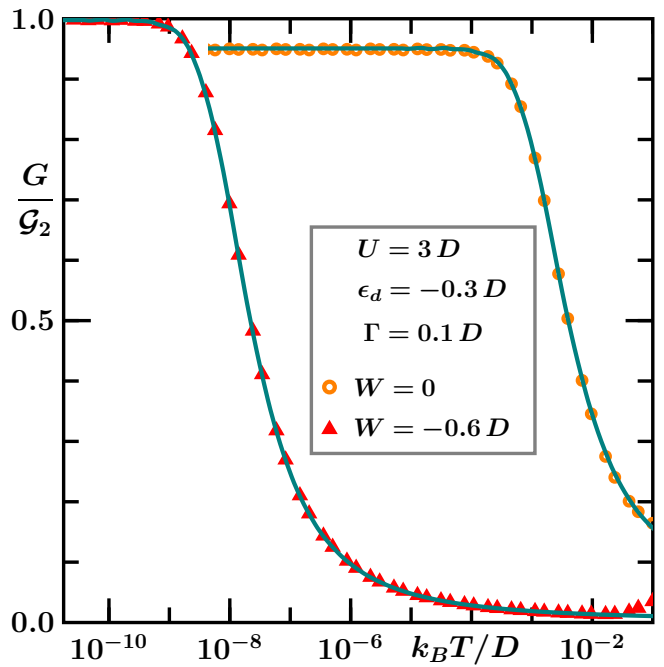}
  \caption{Numerical data for the temperature dependence of the
    conductance, compared to Eq.~(\ref{eq:guniversal}). The open
    circles and triangles show the NRG computed conductances for the
    indicated model parameters. The solid lines represent the mapping
    , with ground-phase shifts calculated from the FL eigenvalues of
    the model Hamiltonian and $T_K$ determined by the condition
    $G(T_K)=\gc/2$. The small disagreement between the triangles and
    the solid line above $k_BT=10^{-2}\,D$ is due to the relatively
    large irrelevant operators introduced by the scattering potential
    $W$, whose contribution to $G$ decays in proportion to $k_BT/D$.}
  \label{fig:6}
\end{figure}

Figure~\ref{fig:6} displays the results of two NRG runs for the same
assymetric parameters $U=3\,D$, $\epsilon=-0.3\,D$, $\Gamma=0.1\,D$,
with two scattering potentials $W=0$ and $W=-0.6\,D$. The open circles
reproduce the dashed curve in Fig.~\ref{fig:4}. The particle-hole
asymmetry, combined with the relatively high ratio between the dot
width and excitation energy, $\Gamma/|\epsilon|=1/3$, place the model
Hamiltonian close to the border of the Kondo domain. Although $W=0$,
the ground-state phase shift deviates significantly from $\pi/2$: from
the FL eigenvalues generated by the NRG diagonalization of the model
Hamiltonian, we find $\delta=0.43\pi$. The solid curve through the
center of the circles is a plot of Eq.~(\ref{eq:guniversal}) with
$\delta-\delta_W=0.43\pi$ and $k_BT_K=4\times10^{-3}\,D$.

The scattering potential $W=-0.6\,D$ reduces the Kondo temperature and
raises the FL conductance $G(T=0)$. The former shrinks to
$k_BT_K=2.2\times10^{-8}\,D$, while the latter rises to nearly
$\gc$. Both changes are due to the reduced dot width $\gammaW =
\cos\delta_W \Gamma$. Here Eq.~(\ref{eq:gammaW}) yields
$\rho\gammaW=0.74\rho\Gamma=0.074$, and the resulting smaller
antiferromagnetic coupling~(\ref{eq:jk}) brings the Kondo
temperature~(\ref{eq:tk}) down five orders of magnitude.

The diminished Kondo temperature indicates that the scattering
potential has pushed the model Hamiltonian deeper inside the Kondo
regime. Other indications are the minute high-temperature conductance;
the nearly ballistic low-temperature conductance; and the
overall similarity between $G(T/T_K)$ and the solid line in
Fig.~\ref{fig:4}.

\subsection{Discussion\label{sec:discussion}}
In the Kondo regime, Eqs.~(\ref{eq:glm})~and (\ref{eq:gfl}) fix the
high- and the low-temperature conductances,
respectively. Equation~(\ref{eq:guniversal}) shows that the universal
function $\gset(T/T_K)$ controls the monotonic transition between the
two limits. For $W=0$, in particular, the fixed-point values depend
only on the ground-state phase shift $\delta$ and are symmetric with
respect to $\gc/2$: $G_{LM}=\gc \cos^2\delta$ and
$G_{FL}=\gc\sin^2\delta$. Thus, depending on $\delta$, the transition from
$G_{LM}$ to $G_{FL}$ can be steeper or flatter. Since $\delta$ can
never depart much from $\pi/2$ in the Kondo regime, the argument of
the trigonometric function on the right-hand side of
Eq.~(\ref{eq:guniversal}) can never depart substantially from $\pi$,
and as indicated by the two curves in Fig.~\ref{fig:4},
$G(T/T_K)\approx\gset(T/T_K)\pm20\%$. By contrast with this crude
estimate, the mapping~(\ref{eq:guniversal}) gives excellent agreement
with the circles in Fig.~\ref{fig:6}.

The wire potential $W$ narrows the dot level and displaces the
ground-state phase shift. Depending on the sign and magnitude of $W$,
the phase shift can take any value in its domain of definition
$-\pi/2\le \delta\le\pi/2$. In the Kondo regime, the Friedel sum rule
nonetheless prevents the difference $\delta-\delta_W$ from straying
away from $\pi/2$. All effects considered, the scattering potential
$W$ displaces the conductance curve towards the symmetric limit
$G(T/T_K)=\gset(T/T_K)$.

These findings are in line with the experimentally established notion
that, in the Kondo regime, SET conductances always decay with
temperature.\cite{GGK+98:5225,GSM+98.156,WFF+00:2105,LSB02:725} A
brief comparison between this behavior and that of the side-coupled
device \cite{KAS+04:035319,SAK+05:066801} seems appropriate. As
demonstrated in Ref.~\onlinecite{SYO2007}, a linear mapping analogous
to Eq.~(\ref{eq:guniversal}) can be established between the
side-coupled conductance and $\gset(T/T_K)$; in that
case, however, the coefficient relating the two functions is
independent of $\delta_W$ and hence free from the constraint imposed
by the Friedel sum rule. Under a sufficiently strong wire potential,
its sign can be reversed. Thus, the thermal dependence of $G_{SC}$ is
tunable:\cite{Kat07:233201} a wire potential can turn a monotonically
increasing function into a monotonically decreasing one. The embedded
geometry of Fig.~\ref{fig:1} is much less sensitive to $W$.

The parameter $\delta_W$ is ($\pi$ times) the charge induced under the
wire electrodes by the potential $W$. According to the Friedel sum
rule, \cite{La66:516} the difference $\delta-\delta_W$ is the charge
of the Kondo cloud, the additional charge that piles up at the
wire tips surrounding the dot as the temperature is lowered past
$T_K$. Neutrality makes the charge of the Kondo cloud equal to the dot
occupancy. Since the symmetric condition $n_d=1$ maximizes the the
low-temperature conductance, one expects $G(T=0)$ to be ballistic for
$2(\delta-\delta_W)=\pi$, a conclusion in agreement with
Eq.~(\ref{eq:guniversal}). Since the screening charge is always nearly
unitary, one expects the low-temperature conductance to be close to the
conductance quantum, in agreement with the plots in Fig.~\ref{fig:6}.

\section{Conclusions\label{sec:conclusions}}

Our central result, Eq.~(\ref{eq:guniversal}) maps the conductance in
the embedded geometry onto the universal conductance for the symmetric
Anderson model~(\ref{eq:hasym}). Different from the universal result
for the side-coupled geometry, the mapping depends explicitly on the
potential $W$ applied to the wire. Section~\ref{sec:cross} showed
that, in the Kondo regime, the Friedel sum rule anchors the
the argument of the cosine on the right-hand side of
Eq.~(\ref{eq:guniversal}) to the vicinity of $\pi$; it results that
$G(T/T_K)$ reproduces semiquantitatively the universal function $\gset(T/T_K)$.

At the quantitative level, Eq.~(\ref{eq:guniversal}) affords
comparison with experimental data collected anywhere in the Kondo
regime. For that purpose, its linearity is particularly
convenient. Once fitted to a set of experimental points, the mapping
determines the Kondo temperature $T_K$, as well as the phase shift
difference $\delta-\delta_W$. Both are quantities of physical
significance. According to the Friedel sum rule, the phase shift
difference is ($\pi/2$ times) the screening charge surrounding the dot
at low temperatures.

In summary, we have derived an exact expression relating the SET
conductance in the Kondo regime to the universal conductance function
for the symmetric Anderson Hamiltonian. A subsequent report will
exploit this result in an attempt to offer a unified view of an NRG
survey of conductance in the Kondo regime.

\acknowledgments

This work was supported by the CNPq and FAPESP.

\appendix
\section{Properties of the fixed-point Hamiltonians\label{sec:diag}}
\subsection{Diagonalization\label{sec:diagonalization}}
The LM and FL are described by conduction-band Hamiltonians of the form
\begin{equation}
  \label{eq:fpHamilt}
H^*= \sum_k \epsilon_k a_k^\dagger a_k + W^*f_0^\dagger f_0.
\end{equation}

We want to bring $H^*$ to the diagonal form
\begin{equation}
  \label{eq:diagFPHamilt}
  H^* = \sum_\ell \varepsilon_\ell g_\ell^\dagger g_\ell,
\end{equation}
where
\begin{equation}
  \label{eq:gks}
  g_\ell = \sum_q \alpha_{\ell q}a_q.
\end{equation}
To this end, we compare the expressions for the commutator
$[g_\ell,H^*]$ obtained from Eqs.~(\ref{eq:fpHamilt})~and
(\ref{eq:diagFPHamilt}), from which it follows that 
\begin{equation}\label{eq:alphaks}
\alpha_{\ell q} = \frac{1}{\varepsilon_\ell -\epsilon_q}\frac{W^*}{N}\sum_k
\alpha_{\ell k}.
\end{equation}

Summation of both sides over $q$ then leads to the eigenvalue condition:
\begin{equation}
  \label{eq:eigenval}
  1 = \frac{W^*}{N} \sum_q\frac{1}{\varepsilon_\ell -\epsilon_q}.
\end{equation}
Inspection of this equality shows that, with exception of a split-off
energy, which makes $\mathcal{O}(1/N)$ contributions to the low-energy
properties, the $\varepsilon_\ell$ are shifted by less than $\Delta$
from the $\epsilon_k$. We therefore refer to the closest conduction
energy $\epsilon_\ell$ to label each eigenvalue and define its phase
shift $\delta_\ell$ with the expression
\begin{equation}\label{eq:delta}
  \varepsilon_\ell \equiv\epsilon_\ell-\frac{\Delta}{\pi}\delta_\ell.
\end{equation}

This definition substituted in Eq.~(\ref{eq:eigenval}), a
Sommerfeld-Watson transformation\cite{matthews71:_mathem_method_of_physic} determines the sum on
the right-hand side:
\begin{equation}
  \label{eq:sommer-wat}
\frac1N\sum_q\frac{1}{\varepsilon_\ell -\epsilon_q} = -\pi\rho \cot
\delta_\ell +\rho \,\fint_{-D}^D\frac{1}{\varepsilon_\ell -\epsilon}\,d\epsilon.
\end{equation}
Substitution in Eq.~(\ref{eq:eigenval}) results in an expression for
the phase shifts:
\begin{equation}\label{eq:cotdelta}
\cot\delta_\ell = -\frac1{\pi\rho W^*} 
+\frac1{\pi} \,\fint_{-D}^D\frac{1}{\varepsilon_\ell -\epsilon}\,d\epsilon.
\end{equation}
At low energies, the contribution of the last term on the right-hand
side, of $\mathcal{O}(\epsilon/D)$, can be neglected, which shows that
the phase shift becomes uniform:
\begin{equation}\label{eq:tandelta}
  \tan\delta= -\pi\rho W^*.
\end{equation}

To determine the coefficients $\alpha_{\ell q}$, we square both sides
of Eq.~(\ref{eq:alphaks}) and sum the result over $q$:
\begin{equation}\label{eq:alpha2}
\sum_q\alpha^2_{\ell q} = \lp\sum_k\alpha_{\ell k}\frac{W^*}{N}\rp^2
\sum_q\frac{1}{(\varepsilon_\ell -\epsilon_q)^2}. 
\end{equation}
To evaluate the sum on the left-hand side, we differentiate
Eq.~(\ref{eq:sommer-wat}) with respect to $\epsilon_\ell$, which
yields, with relative error $\mathcal{O}(1/N)$,
\begin{equation}\label{eq:sumeps2}
\frac1{N^2}\sum_q\frac{1}{(\varepsilon_\ell -\epsilon_q)^2} =
\lp\frac{\pi\rho}{\sin \delta_\ell}\rp^2.
\end{equation}

The sum on the left-hand side of Eq.~(\ref{eq:alpha2}) being unitary,
Eq.~(\ref{eq:sumeps2}) shows that
\begin{equation}\label{eq:eigenvect}
  W^*\sum_k\alpha_{\ell k} = -\frac1{\pi\rho} \sin\delta_\ell,
\end{equation}
the negative phase insuring that $\alpha_{k k}\to1$ for
$W^*\to0$. Equation~(\ref{eq:alphaks}) then gives
\begin{equation}\label{eq:alphasergs}
  \alpha_{\ell q} = \frac{\Delta}{\epsilon_q-\varepsilon_\ell}
  \frac{\sin\delta_\ell}{\pi}\qquad(\varepsilon_\ell\ll D).
\end{equation}

\subsection{Energy moments of the matrix elements of the
  eigenoperators $g_\ell$}
\label{sec:higher}
The Hamiltonian~(\ref{eq:fpHamilt}) diagonalized, we turn our
attention to the following energy moments
\begin{equation}
  \label{eq:moments}
  \mmn{p}\equiv\frac1{\sqrt{N}}\sum_\ell\lp\frac{\varepsilon_\ell}{D}\rp^p\matel{m}{
    g_\ell}{n}\qquad(p=0,1,\ldots).
\end{equation}
Chiefly important are $\mmn{0}\equiv\matel{m}{\phi_0}{n}$ and
$\mmn{1}\equiv\matel{m}{\phi_1}{n}/\sqrt3$; the other moments, as
shown below, are proportional to $\mmn{1}\equiv\matel{m}{\phi_1}{n}$.
Since the $M_{mn}^p$ are universal, to evaluate them it is sufficient
to consider the symmetric Hamiltonian~(\ref{eq:hasym}), for which the
phase shift $\delta_{LM}=0$, so that $g_k$, $\varepsilon_k$, $\phi_0$, and
$\phi_1$ coincide with $a_k$, $\epsilon_k$, $f_0$, and $f_1$, respectively.

From Eq.~(\ref{eq:hasym}), we then have that
\begin{equation}
  \label{eq:syscomm}
  [g_\ell,\ha^S]   = \varepsilon_\ell g_{\ell} + \frac V{\sqrt N}\cd.
\end{equation}
With the shortand $\emn\equiv(E_m-E_n)/D$, the multiplication of both
sides by $(\varepsilon_\ell/D)^{p-1}$ followed by summation over
$\ell$ leads to the coupled recursive relations
\begin{equation*}
  \begin{array}{lll}
  \mmn{p} =& -\emn\mmn{p-1}-
  \displaystyle\frac V{p}\matel{m}{\cd}{n}
  &\qquad(p=1,3,\ldots);\\
  \mmn{p}=&-\emn\mmn{p-1}&\qquad(p=2,4,\ldots).
\end{array}
\end{equation*}
Reduced to a matrix equation, this system is easily solved. The result is
\begin{equation}
\mmn{p} =\left\{\begin{array}{ll}
 \mmn{0}\emn^p-\displaystyle\frac Vp\matel{m}{c_d}{n}\lp1+
{\displaystyle\sum_{r=1}^{p-1}}
\mbox{\raisebox{7pt}{$\prime$}}\frac{{\emn}^r}{r}\rp&\quad(p=\mathrm{odd})\nonumber\\
 \mmn{0}\emn^p-\displaystyle\frac Vp\matel{m}{c_d}{n}
{\displaystyle\sum_{r=1}^{p-1}}
\mbox{\raisebox{7pt}{$\prime$}}\frac{{\emn}^r}{r}&\quad(p=\mathrm{even})\nonumber
\label{eq:matrixmv}
\end{array}\right.,
\end{equation}
where the primed sum is  restricted to odd $r$'s.

The pertinent energies satisfy the condition $k_BT\ll D$, which implies
$\emn \ll 1$.  It is therefore safe to discard the terms proportional
to $\emn$ and its powers. Within this approximation, the only nonzero
even moment is $\mmn{0}$, and all odd moments are proportional to
$\matel{m}{\cd}{n}$. It follows that all the odd moments are
proportional to $\mmn{1}=\matel{m}{\phi_1}{n}$:
\begin{equation}
    \label{eq:mmnsvsmmn1}
    \mmn{p}=\left\{
      \begin{array}{ll}
        \displaystyle\frac{\matel{m}{\phi_1}{n}}p&\qquad(p=\text{odd})\\
        0&\qquad(p=2,4,\ldots)
      \end{array}\right..
\end{equation}

An orthonormal basis describing the conduction
band~(\ref{eq:diagFPHamilt}) can be constructed from the definition
\begin{equation}
  \label{eq:lagrange}
  \phi_p \equiv
  \sqrt{\frac{2p+1}N}\sum_{\ell}P_p(\epsilon_\ell)g_{\ell}
  \qquad(p=0,1,\ldots),
\end{equation}
where $P_p(\epsilon)$ denotes a Legendre polynomial.

According to Eq.~(\ref{eq:mmnsvsmmn1}), 
$\matel{m}{\phi_p}{n}\sim\matel{m}{\phi_1}{n}$ ($p=3,5,\ldots$), while
$\matel{m}{\phi_p}{n}=0$ ($p=2,4,\ldots$). This shows that the matrix
element of any conduction operator is a linear combination of
$\matel{m}{\phi_0}{n}$ and $\matel{m}{\phi_1}{n}$. In particular
\begin{equation}
  \label{eq:f0matelphi0}
  \matel{m}{f_i}{n} = \sum_{j=0}^1\alpha_{ij}\matel{m}{\phi_j}{n}\qquad(i=0, 1),
\end{equation}
where $f_0$ and $f_1$ are the operators defined by
Eqs.~(\ref{eq:f0})~and (\ref{eq:f1}), respectively, and the
$\alpha_{ij}$ ($i,j=0,1$) are constants that depend on the model parameters.

\section{Fixed-point conductances\label{sec:extens-langr}}
This appendix derives an expression for the spectral density
$\rho_d(\epsilon,T)$, defined by Eq.~(\ref{eq:rhod}), at the fixed
points. The procedure is analogous to the one in
Appendix~\ref{sec:diag}.

From Eq.~(\ref{eq:ha_comm_aq}) we obtain an expression for the matrix
element of $a_q$ between two low-energy eigenstates $\ket m$ and $\ket
n$ of eigenstates of $\ha$:
\begin{eqnarray}\label{eq:amatel}
  \matel{m}{a_q^\dagger}{n} &=& \frac1{\sqrt{N}}\frac{V}{E_m-E_n-\epsilon_q}
  \matel{m}{\cdd}{n}\nonumber\\
  &&+\frac W{N}\frac{1}{E_m-E_n-\epsilon_q}
  \matel{m}{\sum_p a_p^\dagger}{n}.
\end{eqnarray}
Summation of both sides over $q$ leads to an expression for the matrix
element in the last term on the right-hand side:
\begin{eqnarray}
  \matel{m}{\sum_p a_p^\dagger}{n} \lp1-
  W\smn\rp=\sqrt N V\matel{m}{\cdd}{n}\smn,
\end{eqnarray}
where
\begin{equation}
  \label{eq:smn}
  \smn\equiv\frac1N\sum_q\frac{1}{E_m-E_n-\epsilon_q},
\end{equation}
which brings Eq.~(\ref{eq:amatel}) to the form
\begin{equation}\label{eq:amatelnosum}
  \matel{m}{a_q^\dagger}{n} =
  \frac{\matel{m}{\cdd}{n}}{\sqrt N(E_m-E_n-\epsilon_q)}
  \frac{V}{1-W\smn}.
\end{equation}

Consider now this equality at one of the two fixed points, LM or
FL. The fixed-point Hamiltonian has then the quadratic
form~(\ref{eq:diagFPHamilt}), which defines the complete basis of the
operators $g_\ell$. The matrix element $\matel{m}{g_\ell^\dagger}{n}$
vanishes unless $\ket{m}=g_\ell^\dagger\ket{n}$, which implies
$E_m=E_n+\varepsilon_\ell$. At a fixed point, therefore, the sum on the right-hand side of
Eq.~(\ref{eq:smn}) reduces to that in Eq.~(\ref{eq:sommer-wat}), \ie\
\begin{equation}
  \label{eq:smnFP}
  \smn= -\pi\rho \cot \delta_*,
\end{equation}
where we have disconsidered the last term on the right-hand side of
Eq.~(\ref{eq:sommer-wat}) because at a fixed point the ratio
$\varepsilon_\ell/D\to 0$. This result suggests that we introduce the
phase shift $\delta_W$, defined by
\begin{equation}
  \label{eq:deltaw}
  {W} \equiv -\frac{\tan\delta_W}{\pi\rho},
\end{equation}
to simplify Eq.~(\ref{eq:amatelnosum}):
\begin{equation}
  \label{eq:amatelnosum_simple}
  \matel{m}{a_q^\dagger}{n} =
  \frac{V\matel{m}{\cdd}{n}}{\sqrt N(E_m-E_n-\epsilon_q)}
  \frac{\sin\delta_*\cos\delta_W}{\sin(\delta_*-\delta_W)}.
\end{equation}
In analogy with Eq.~(\ref{eq:gks}) we can, moreover, write
\begin{equation}
  \label{eq:gkAlphasCd}
  g_\ell = \gamma_{\ell0} \cd + \sum_q\gamma_{\ell q}a_q,
\end{equation}
with normalized coefficients:
\begin{equation}
  \label{eq:norm_gks}
  \gamma_{\ell 0}^2+\sum_q \gamma_{\ell q}^2 = 1.
\end{equation}

At each fixed point, therefore, once squared, Eq.~(\ref{eq:amatelnosum}) reads
\begin{equation}
  \label{eq:aq_cd_sum}
    \gamma_{\ell q}^2=
  \frac{V^2\gamma_{\ell 0}^2}{N(\varepsilon_\ell-\epsilon_q)^2}
  \lp\frac{\sin\delta_*\cos\delta_W}{\sin(\delta_*-\delta_W)} \rp^2.
\end{equation}
We divide both sides by $N$, sum them over $q$, and substitute
Eq.~(\ref{eq:sumeps2}) for the resulting sum on the right-hand side, to find that
\begin{equation}
  \label{eq:aq_cd_delta}
    \sum_q\gamma_{\ell q}^2= {NV^2}{\gamma_{\ell 0}^2}
    \lp\frac{\pi\rho\cos\delta_W}{\sin(\delta_*-\delta_W)} \rp^2.
\end{equation}

Substitution in the second term on the left-hand side of
Eq.~(\ref{eq:norm_gks}) now shows that, with error $\mathcal{O}(1/N)$:
\begin{equation}
  \label{eq:gamma_l0}
    |\matel{m}{\cdd}{n}|^2= \frac1{NV^2}
  \frac{\sin^2(\delta_*-\delta_W)}{\pi^2\rho^2\cos^2\delta_W}.
\end{equation}

A the fixed points, the matrix elements are constants, dependent only
on the phase shift and scattering potential. The spectral density
$\rho_d$, as one would expect, becomes independent of the temperature:
\begin{equation}
  \rho_{d}(\epsilon,T)=\frac1{NV^2{\zpart}}
  \sum_{m,n} e^{-\beta E_m}
  \frac{\sin^2(\delta_*-\delta_W)}
  {\pi^2\rho^2\cos^2\delta_W}\delta(\epsilon_\ell-\epsilon),
\end{equation}
equivalent to
\begin{equation}
  \label{eq:rho_d}
  \rho_{d}(\epsilon)=\frac{\sin^2(\delta_*-\delta_W)}
  {\pi\Gamma\cos^2\delta_W}.
\end{equation}

$W=0$ recovers the celebrated expression\cite{La66:516}
\begin{equation}
  \label{eq:langreth}
  \rho_{d}(\epsilon)=\frac{\sin^2\delta_*}
  {\pi\Gamma}.
\end{equation}
More generally, however, to obtain the fixed-point spectral densities,
we set $\delta^*=\delta$ at the FL, and $\delta^*=\delta-\pi/2$ at the
LM, from which it results that
\begin{subequations}
\label{eq:rho_FPs}
  \begin{eqnarray}    
    \label{eq:rho_LM}
    \rho_d^{LM}&=&
    \frac{\cos^2(\delta-\delta_W)}{\pi\Gamma\cos^2\delta_W};\\
    \rho_d^{FL}&=&
    \frac{\sin^2(\delta-\delta_W)}{\pi\Gamma\cos^2\delta_W}.
    \label{eq:rho_FL}
  \end{eqnarray}
\end{subequations}
Substitution of Eqs.~(\ref{eq:rho_LM})~and~(\ref{eq:rho_FL}) for
$\rho_d$ on the right-hand side of Eq.~(\ref{eq:glin}) leads to
Eqs.~(\ref{eq:glm})~and (\ref{eq:gfl}), respectively.

\section{Zero-bias conductance\label{sec:expr-cond}}
By contrast with the coupling to the impurity, which is independent of
the odd operators $b_k$ defined by Eq.~(\ref{eq:bk}), the Hamiltonian
describing a bias voltage couples to the $b_k$'s. Preliminary to the
discussion of the conductance, it is therefore convenient to derive
results for the $b_k$'s analogous to those in
Appendix~\ref{sec:diag}. Specifically, given the formal equivalence
between Eqs.~(\ref{eq:hb})~and (\ref{eq:fpHamilt}), we can follow the
steps in that Appendix to write $H_B$ in the diagonal form
\begin{equation}
  \label{eq:hbdiag}
  H_B=\sum \tilde\varepsilon_\ell {\tilde g}^\dagger_\ell\tilde g_\ell,
\end{equation}
with
\begin{equation}
  \label{eq:glbetakl}
  \tilde g_\ell = \sum_k \tilde \alpha_{\ell,k} b_k,
\end{equation}
and to derive a result analogous to Eq~(\ref{eq:eigenvect}). At low
energies, in particular, \ie\ for $|\tilde\varepsilon_\ell|\ll D$, the
eigenvalues $\tilde\varepsilon_\ell$ are uniformly phase shifted by
$\delta_W$ [see Eq.~(\ref{eq:deltaw})], and
\begin{equation}
\sum_k \tilde\alpha_{\ell,k} = \cos\delta_W\label{eq:sumtildalphas}.
\end{equation}
Multiplication of both sides by $\tilde g_\ell$ and summation over
$\ell$ then shows that
\begin{equation}
  \label{eq:gl_matell}
  \sum_k\matel{\tilde m}{b_k}{\tilde n} =\cos\delta_w \sum_\ell\matel{\tilde
    m}{\tilde g_\ell}{\tilde n},
\end{equation}
for any pair $\ket{\tilde m},\ket{\tilde n}$ of low-energy
eigenstates of $H_B$.

It is likewise convenient to compute the following commutator:
\begin{equation}\label{eq:abmatel_raw}
  [\ha,a_k^\dagger b_k] = \frac{V}{\sqrt{N}} \cdd\, b_k
+\frac{W}{N}\sum_q(a_q^\dagger b_k- a_k^\dagger b_q),
\end{equation}
from which we see that, given two eigenstates $\psim$ and
$\ket{n}$ of $\ha$ with eigenvalues $E_m$ and $E_n$, respectively,
\begin{eqnarray}\label{eq:abmatel}
  \psimatel{a_k^\dagger b_k}{n} &=&
\frac{V}{\sqrt{N}}\frac{\psimatel{\cdd\,b_k}{n}}{E_m-E_n}\nonumber\\
&&+\frac{W}{N}\sum_q\frac{\psimatel{a_q^\dagger b_k- a_k^\dagger b_q}{n}}{E_m-E_n}.
\end{eqnarray}

 \subsection{Current\label{sec:current}}
To calculate the conductance, we can, for instance, examine the
current flowing into the $R$ wire:
\begin{equation}
  \hat I = \frac{dq_R}{dt}=-\frac{i\ee}{\hbar}
       [\ha,\sum_{k} c_{kR}^\dagger c_{kR}],
\end{equation}
\ie\ 
\begin{equation}\label{eq:current}
  \hat I =-\frac{i\ee}{2\hbar}
  [\ha,\sum_{k}\lp a_{k}^\dagger a_{k}+ b_{k}^\dagger b_{k} 
  -(a_{k}^\dagger b_{k}+\hc)\rp],
\end{equation}
which is equivalent to
\begin{equation}
  \label{eq:currentNoW}
  \hat I = \frac{i\ee}{2\hbar}\frac{V}{\sqrt N}\,\cdd\sum_{k}(a_k+b_k)+\hc,
\end{equation}
because summed over $k$, the last term on the right-hand side of
Eq.~(\ref{eq:abmatel_raw}) vanishes.
\subsection{Conductance\label{sec:conductance}}
To induce a current, we add to the model Hamiltonian an infinitesimal,
slowly growing perturbation that lowers the chemical potential of the
$R$ wire relative to that of the $L$ wire:
\begin{equation}
  H_\mu \equiv \Delta\mu\,\hnu(t) = -\ee\frac{\Delta\mu}2 \sum_{k}\lp c_{kR}^\dagger
  c_{kR}-c_{kL}^\dagger c_{kL}\rp e^{\eta t/\hbar},
\end{equation}
with an infinitesimal shift $\Delta\mu$. 

Projected on the basis of the $a_k$'s and $b_k$'s, $\hnu$ reads
\begin{equation}\label{eq:hnu}
  \hnu(t) = -\frac{\ee}2\sum_{k}(a_{k}^\dagger b_{k}+\hc)e^{\eta t/\hbar},
\end{equation}
and Eq.~(\ref{eq:abmatel}) shows that
\begin{equation}\label{eq:hnumatel}
  \psimatel{\hnu(t)}{n} =-\frac{\ee V}{2\sqrt{N}}e^{\eta t/\hbar}\sum_{k>0}
  \frac{\psimatel{\cdd b_k- b_k^\dagger\cd}{n}}{E_m-E_n}.
\end{equation}

Standard Linear Response Theory relates $\hnu$ to the conductance:
\begin{equation}\label{eq:gih}
  G(T) = -\frac{i}{\zpart\hbar}
  \int_{-\infty}^{0}\sum_m e^{-\beta E_m}\psimatel{[\hat I,\hnu(t)]}{m}\,dt,
\end{equation}
where $\zpart$ is the partition function at the temperature $T$.

Comparison with Eq.~(\ref{eq:hnu}) shows that the operators $a_k$
within the parentheses on the right-hand side of that equality make
no contribution to the conductance. We therefore define
\begin{equation}\label{eq:ib}
  \hat I_b \equiv \frac{i\ee}{2\hbar}\frac{V}{\sqrt N}\,\cdd\sum_k b_k +\hc,
\end{equation}
and rewrite Eq.~(\ref{eq:gih}):
\begin{equation}\label{eq:gcom}
  G(T) = -\frac{i}{\zpart\hbar}
  \int_{-\infty}^{0}\sum_m e^{-\beta E_m}\psimatel{[\hat I_b,\hnu(t)]}{m}\,dt.
\end{equation}

Following the insertion of a completeness sum $\sum_n\ket{n}\bra{n}$ on the
right-hand side of Eq.~(\ref{eq:gcom}), straightforward
manipulations lead to the familiar expression
\begin{equation*}
  G(T) = \frac{1}{\zpart}
  \sum_{m,n} \lp e^{-\beta E_m}-e^{-\beta E_n}\rp
  \frac{\psimatel{\hat I_b}{n}\psinmatel{\hnu(0)}{m}}{E_m-E_n+i\eta}.
\end{equation*}

On the right-hand side, we now substitute Eq.~(\ref{eq:hnumatel})
[Eq.~(\ref{eq:ib})] for $\hnu$ ($\hat I_b$). This yields
  \begin{eqnarray}\displaystyle
    G(T) = -i\frac{\ee^2}{4\hbar}\frac{V^2}{N\zpart} \sum_{m,n,k,q}
    &\displaystyle\lp\frac{\psimatel{b_{q}^\dagger
        \cd}{n}\psinmatel{\cdd b_{k}}{m}}{E_m-E_n+i\eta}\right.\nonumber\\
    &\left.\displaystyle+\frac{\psimatel{\cdd b_k}{n}
        \psinmatel{b_{q}^\dagger \cd}{m}}
    {E_m-E_n+i\eta}\rp\nonumber\\
    &\times\displaystyle\frac{e^{-\beta E_m}-e^{-\beta E_n}}{E_m-E_n}.
  \end{eqnarray}
  Aided by Eq.~(\ref{eq:gl_matell}), we can now trade the sum over the
  conduction operators $b_k$ for a sum over the eigenoperators
  $\tilde g_{\ell}$:
\begin{subequations}
  \begin{eqnarray}\displaystyle
    G(T) = -i\frac{\ee^2}{2h}\frac{\gammaW}{N\rho\zpart} 
    \sum_{m,n,\ell,\ell'}&&
    \displaystyle\lp\frac{\psimatel{\tilde g_{\ell'}^\dagger
        \cd}{n}\psinmatel{\cdd \tilde g_{\ell}}{m}}{E_m-E_n+i\eta}\right.\nonumber\\
    &&\left.\displaystyle+\frac{\psimatel{\cdd
          \tilde g_\ell}{n}\psinmatel{\tilde g_{\ell'}^\dagger \cd}{m}}
    {E_m-E_n+i\eta}\rp\nonumber\\
&\times&\frac{e^{-\beta E_m}-e^{-\beta E_n}}{E_m-E_n}.
  \end{eqnarray}
\end{subequations}
Since the $\tilde g_\ell$ diagonalize $H_B$, only the terms with
$\ell=\ell'$ contribute to the sum on the right-hand side. We
interchange the indices $m$ and $n$ in the second term within the
parentheses on the right-hand side to show that
\begin{equation}\label{eq:gdelta}
G(T) = \frac{\pi\ee^2}{h}\frac{\beta\,\Gamma_w}{\rho N\zpart} \sum_{m,n,\ell}e^{-\beta E_m}
|\psimatel{\cdd g_{\ell}}{n}|^2\delta(E_m-E_n).
\end{equation}

Since $\psim=\ket{m}\ket{\tilde m}$, where $\ket{m}$ ($\ket{\tilde
  m}$) is an eigenstate of $H_A$ (of the quadratic Hamiltonian $H_B$),
the right-hand side splits into two coupled sums:
\begin{eqnarray}\label{eq:oneboltzmann}
  G(T) =\frac{\pi\ee^2}{h}\frac{\beta\,\gammaW}{N\rho\zpart}
  &\displaystyle\sum_{m,n,\ell}e^{-\beta E_{m}}
  |\matel{m}{V\cdd}{n}|^2\delta(E_{m}-E_{n}-\tilde\epsilon_\ell)\nonumber\\
  &\displaystyle\times\sum_{\tilde m,\tilde n}e^{-\beta E_{\tilde m}}\matel{\tilde
    m}{g_{\ell}}{\tilde n}\matel{\tilde n}{g_\ell^\dagger}{\tilde m}.
\end{eqnarray}

The second sum is equal to $\zpart_b[1-f(\tilde\epsilon_p)]$, where
$f(\epsilon) $ is the Fermi function,  and $\zpart_b$, the partition
function for the Hamiltonian $H_b$. The identity 
\begin{equation}
  -\frac1{f(\epsilon)}\frac{\partial f}{\partial \epsilon}= \beta \lp1-f(\epsilon)\rp
\end{equation}
then turns Eq.~(\ref{eq:oneboltzmann}) into
\begin{eqnarray}\label{eq:gforha}
  G(T) = \frac{\ee^2}{h\zpart_a}\frac{\pi\gammaW}{\rho N}
  \sum_{m,n,\ell}&\displaystyle\frac{e^{-\beta E_{m}}}{f(\tilde\epsilon_\ell)} 
  \lp-\frac{\partial f}{\partial \epsilon}\rp_{\tilde\varepsilon_\ell}  
  |\matel{m}{\cdd}{n}|^2\nonumber\\&\times\,\delta(E_{m}-E_{n}-\tilde\epsilon_\ell),
\end{eqnarray}
where $\zpart_a$ is the partition function for the Hamiltonian $\ha$.

The definition (\ref{eq:rhod}) of the spectral density
$\rho_d(\epsilon, T)$ allows us to rewrite Eq.~(\ref{eq:gforha}) as
\begin{equation}
  G(T) = \frac{\ee^2}{h}\frac{\pi\gammaW}{\rho N}
  \sum_{\ell} \lp-\frac{\partial f}{\partial
    \epsilon}\rp_{\tilde\varepsilon_\ell}\rho_{d}(\tilde\epsilon_\ell),
\end{equation}
from which Eq.~(\ref{eq:glin}) follows.

\begin{thebibliography}{42}
\bibitem[{irr()}]{irrelevant}
\bibinfo{note}{A sum of irrelevant terms such as $\sum_k\epsilon
  c_{k\mu}^\dagger \bm{\sigma}_{\mu\nu}\phi_{0\nu}\bm{S}$ or $\sum_k\epsilon_k
  c_{k\mu}^\dagger \bm{\sigma}_{\mu\nu}\sum_k\epsilon_k c_{k\nu}\bm{S}$, whose
  contribution to physical properties is $\mathcal{O}(k_BT/)\ll 1$, has been
  neglected on the right-hand side of Eq.~(\ref{eq:lmkondo}).}
\end{thebibliography}

\end{document}